\documentclass[twocolumn,showpacs]{revtex4}
\usepackage{amsfonts}
\usepackage{amsmath}
\usepackage{graphicx}

\begin{document}
\draft
\title{Surface superconducting states
in yttrium hexaboride single crystal in a tilted magnetic field}

\author{M.I. Tsindlekht, V.M. Genkin, G.I. Leviev}
\affiliation{The Racah Institute of Physics, The Hebrew
University of Jerusalem, 91904 Jerusalem, Israel}
\author{N.Yu. Shitsevalova}
\affiliation{Institute for Problems of Materials Science, National Academy of
Sciences of Ukraine, 03680 Kiev, Ukraine}

\begin{abstract}
We present the results of an experimental study of the nucleation of superconductivity at the surface of a single crystal YB$_6$ in a tilted dc magnetic field. A recently developed experimental technique allowed us to determine $H_{c3}$ at each side of the sample as a function of the angle between the dc magnetic field and the surface. Experiment shows that the ratio $H_{c3}/ H_{c2}\approx 1.28 $ in the direction perpendicular to the surface dc field while according to the theory this ratio should be equal to 1. This sharp distinction cannot be ascribed to the surface roughness.
\end{abstract}

\date{\today}
\maketitle

\section{Introduction}
Shortly after Saint-James and de Gennes' prediction of surface superconductivity~\cite{PG} many experiments have confirmed their basic idea. Resistive and permeability measurements at low frequencies~\cite{KIM,STR,TOM,BURG,ROLL} showed that in parallel to the surface dc field the ratio $H_{c3}/H_{c2}$ is close to the theoretical value 1.695. The value of $H_{c3}$ depended on the angle $\theta$ between the dc field and the sample surface and decreased as this angle was increased. The critical surface current, defined as the current at which the sample exhibits a finite resistance, also decreased with increasing $\theta$. The ac response of surface superconducting states (SSS) had characteristic features such as the loss maximum in the intermediate fields $H_{c2}<H_0<H_{c3}$, low frequency dispersion, and nonlinearity at very low excitation levels~\cite{ROLL}. Concurrently, a number of theoretical models for the calculation of the critical current in the SSS were published (see, for example, references in~\cite{ROLL}). Agreement between the theoretical models and experimental data was not satisfactory. Therefore, attempts were made to develop a more sophisticated model of the SSS, in particular, a model of the surface vortices was elaborated~\cite{SWR,KUL,KAR}. In this model the surface vortices could move only if the surface current exceeds some critical value. Actually this is a variant of the critical state model which was applied to the SSS. When the dc magnetic field is parallel to the sample surface, the normal component is zero and the density of the surface vortices is zero. It was proposed~\cite{SWR} that due to the surface roughness, locally the dc magnetic field had a normal component which provided some finite density of surface vortices. The experiments~\cite{SWR,KAR} demonstrated that the surface critical current dramatically depended on the surface properties. Losses in the critical state model have a threshold character with respect to the amplitude of excitation~\cite{ROLL}. All the above mentioned experiments were performed on polycrystalline metals or alloys. In spite of all these researches an adequate theory that could explain all observed peculiarities has not yet been proposed. In addition, recent experiments did not reveal any vanishing of losses in weak ac fields~\cite{JUR,SHT} and showed that the nonlinear response could not be described by perturbation theory~\cite{SHT}. Resistive and torque measurements~\cite{KIM,BURG,TOM} demonstrated that SSS did not exist ($H_{c3}/H_{c2}=1$) if the dc field was perpendicular to the surface as it was predicted by the theory~\cite{PG}. On the other hand, ac measurements~\cite{STR} demonstrated that in a perpendicular field the transition to a superconducting state takes place in fields above $H_{c2}$. It was assumed that there was a superconducting network of negligible volume that was not detected in any bulk effects~\cite{STR}. The main goal of this paper is to demonstrate that SSS could exist at $H_0 > H_{c2}$ perpendicular to the surface dc field, and this effect could not be ascribed to the surface roughness of the sample.

\section {Experimental details}

\begin{figure}
    \begin{center}
  \leavevmode
\includegraphics[width=0.9\linewidth]{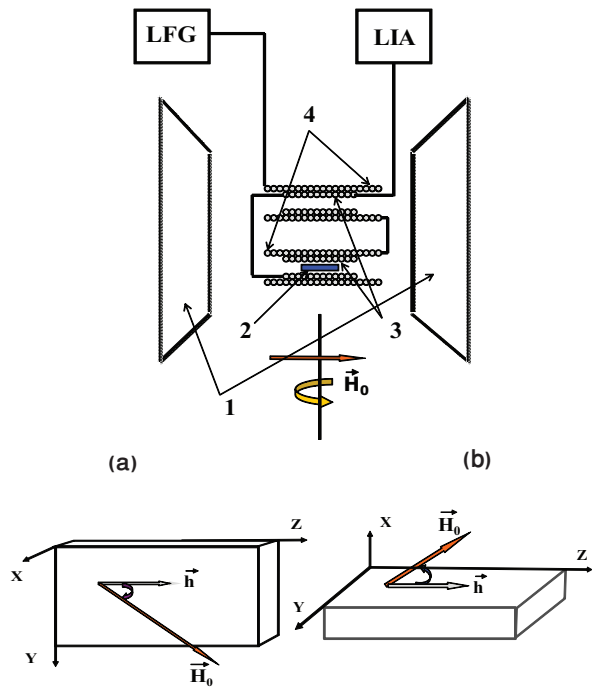}

\caption{(Color online)  Block diagram of the experimental setup. LFG - low frequency generator, LIA - lock-in amplifier, 1 - electromagnet poles, 2 - sample, 3 - pickup coils, and 4 - drive coils. Inset (a) - sketch of the dc field rotation in \textit{XZ}-plane. Inset (b) - sketch of the dc field rotation in \textit{YZ}-plane.}

\label{f-1}
\end{center}
\end{figure}

We have measured the ac response of a YB$_6$ single crystal with sizes 10 by 3 by 1 mm  which was cut from a large crystal. Sample surfaces were mechanically and then chemically polished. DC magnetization curves were measured by a commercial SQUID magnetometer. The ac response at the fundamental frequency and the third harmonic signal were taken concurrently using the pickup coil
method~\cite{SH}. A block diagram of the experimental setup is shown in Fig.~\ref{f-1}. The ac field is supplied by two identical drive coils connected in series with the load resistor. The sample is inserted into one of the pickup coils. Empty pickup coils are balanced without any external circuits. A small unbalanced signal of the empty pickup coils was subtracted as well. The electromagnet was rotated around the vertical axis, Fig.~\ref{f-1}. Small probe coil (not shown in Fig.~\ref{f-1}) is employed for ac amplitude calibration.

Fourier analysis of the time-dependent magnetization in an ac field $h(t)=2h_0\cos(\omega t)$, yields an expression of the form $m(t)=h_{0}\sum_{n}\chi_n(\omega)\exp(-in\omega t)$. The frequency and amplitude of the applied ac field were 565 Hz and  0.05 Oe, respectively. This frequency has been chosen by taking into account our experimental constrains and results of~\cite{SHT} that the frequency dispersion was not important for the managing superconducting transition.
We assume that in zero dc field at low temperatures the losses are negligible and the ac in-phase component, $\chi_1^{\prime}$, does not depend on frequency and equals to $-1/4\pi$. This assumption permits us to get the values of $\chi_1$ and $\chi_3$ in absolute units at finite dc fields. The absence of frequency dispersion in zero dc field was verified experimentally.

Two experimental configurations were chosen for the parallelepiped shaped sample. The first one, (G1), when the dc field rotates in \textit{XZ}-plane (Fig.~\ref{f-1}, inset (a)), and the second one, (G2), when the dc field rotates in \textit{YZ}-plane ( Fig.~\ref{f-1}, inset (b)). With the first configuration, G1, dc field remains parallel to the small faces (\textit{XZ}-plane) for any angle $\theta$ while with G2 the field is always parallel to the wide faces (\textit{YZ}-plane). Angle $\theta$ is the angle between axis $Z$ and direction of the dc magnetic field. In all measurements the ac field was parallel to the axis $Z$. The measured ac response for these two configurations was different and therefore it permitted us to calculated the ac response for each sample face as a function of the dc field inclination angle $\theta$.

\section{Experimental results}

\begin{figure}
     \begin{center}
    \leavevmode
 \includegraphics[width=0.9\linewidth]{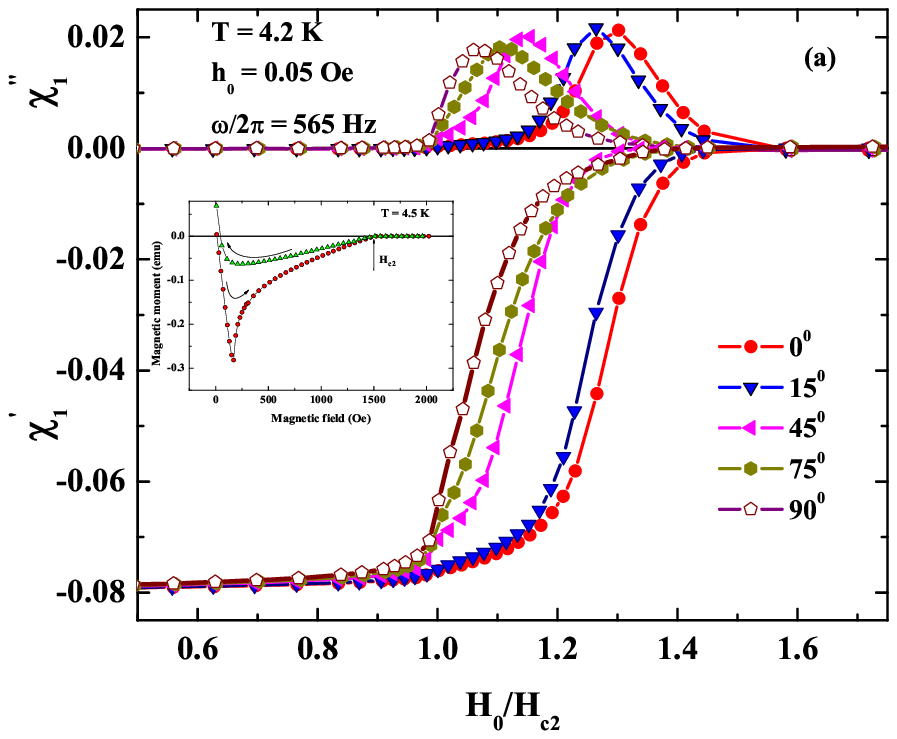}
\includegraphics[width=0.9\linewidth]{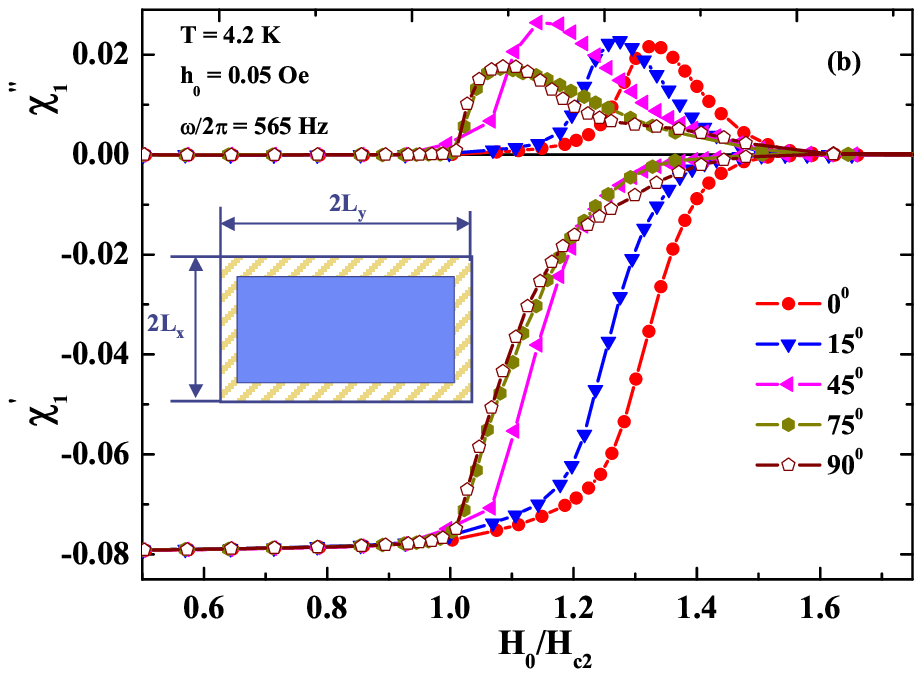}

\caption{(Color online)  Field dependence of $\chi_1^{\prime}$ and
$\chi_1^{\prime\prime}$ for different angles $\theta$. Panel (a): $\chi_1$ for orientation of the dc field that is parallel to the narrow side of the sample and has an angle $\theta$ with wide side (rotation in \textit{XZ}-plane).
Magnetization curve is presented in an upper inset.\\
Panel (b): $\chi_1$ for orientation of the dc field that is parallel to the wide side of the sample and has an angle $\theta$ with narrow side (rotation in \textit{YZ}-plane). The inset shows the sample cross section. The darkened area shows (not in scale) the sample part in a normal, and light colored part in a superconducting state, respectively.}

     \label{f-2}
     \end{center}
\end{figure}

YB$_{6}$ is an isotropic BCS superconductor. It has some advantages due to experimentally well define  $H_{c2}$ as one can see in the inset to Fig.~\ref{f-2}a. Properties of YB$_6$ were measured with high accuracy in Refs.~\cite{JUNO, SHT} where it was found that T$_c\approx~7.2$ K and the Ginzburg-Landau (GL) parameter $\kappa\approx 3.5$.

Fig.~\ref{f-2}a presents in- and out-of-phase components of $\chi_1$ for $T =4.2$ K as a function of the dc field for some angles $\theta$ between the wide sample side (\textit{YZ}-plane) and the dc field while it remained parallel to the narrow side (\textit{XZ}-plane) .
Fig.~\ref{f-2}b shows the corresponding results for another orientation of the dc field when it was parallel to the wide sides and had an angle with the narrow sides (see inset (b) to Figs.~\ref{f-1}). One can readily see that for these two dc field orientations (G1 and G2)  $\chi_1^{\prime}(H_0)$ and $\chi_1^{\prime\prime}(H_0)$ are well distinguished.

\section{Discussion}

\begin{figure}
     \begin{center}
    \leavevmode
 \includegraphics[width=0.9\linewidth]{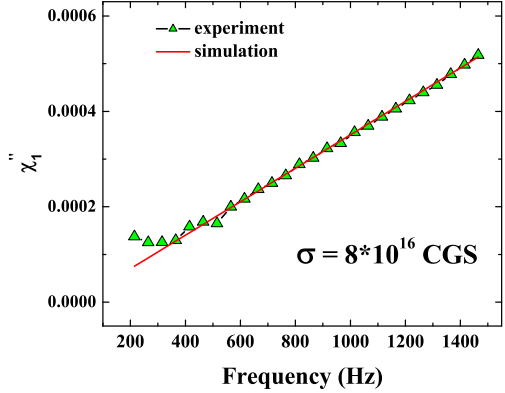}

\caption{(Color online) Frequency dependence of $\chi_1^{\prime\prime}$ for the sample in a normal state at T=7.5 K.}
     \label{f-3a}
     \end{center}
     \end{figure}

The raw experimental data (Fig.~\ref{f-2}a and \ref{f-2}b) contain the contribution of four sample sides to the ac response. In order to separate the response from each side we used the following procedure. Let consider the ac response of the long rectangular slab in the normal state with a thin superconducting sheath (of thickness several coherence length) at the surface (inset to Fig.~\ref{f-2}b). The ac field in the bulk could be found from the solution of the two-dimensional equation
\begin{equation}\label{Eq1}
\frac{\partial^2h_z}{\partial x^2}+\frac{\partial^2h_z}{\partial
y^2}+\frac{2i}{\delta^2}h_z=0,
\end{equation}
where $\delta=c/\sqrt{2\pi\sigma\omega}$ is the normal skin depth and $\sigma$ is a normal conductivity. We took into account that an applied ac field has only one \textit{Z}-component and neglected the small demagnetization factor along the $Z$-axis $\approx 0.045$.
A surface sheath is capable of carrying a surface current $J_s$. This current screens the inner part of the sample and, neglecting the thickness of this sheath, we can write the boundary condition for Eq.(\ref{Eq1}) at each sample side in the form
\begin{equation}\label{Eq2}
h_z^{si}=h_0+4\pi J_{si}/c,
\end{equation}
where $h^{si}$ is the amplitude of ac field on the boundary between normal sample core and the surface sheath at \textit{i-th} sample side, and $J_{si}$ is surface current at \textit{i-th} side.  Indexes $i=1, 2$ correspond to the wide sides while $i=3, 4$ to the narrow sides.
The straightforward solution of Eq. (\ref{Eq1}) is
\begin{eqnarray}\label{Eq3a}
  h_z&=&\sum_{n=1,3...}\frac{4}{\pi n}\sin(\pi n/2)\cos(\pi nx/2L_x)\nonumber\\
  &\times&\left\{\frac{h^{x1}\sinh[k_n(L_y+y)]}{\sinh[2k_nL_y]}
  +\frac{h^{s2}\sinh[k_n(L_y-y)]}{\sinh[2k_nL_y]}\right\}\nonumber\\
  &+&\sum_{n=1,3...}\frac{4}{\pi n}\sin(\pi n/2)\cos(\pi ny/2L_y)\\
 &\times& \left\{\frac{h^{x3}\sinh[q_n(L_x+x)]}{\sinh[2q_nL_x]}
 +\frac{h^{s4}\sinh[q_n(L_x-x)]}{\sinh[2q_nL_x]}\right\},\nonumber
\end{eqnarray}
where $$k_n^2\equiv(\pi n/2L_x)^2-2i/\delta^2,~q_n^2\equiv(\pi n/2L_y)^2-2i/\delta^2,$$ $L_x$ and $L_y$ are sample sizes in \textit{XY}-plane ( $L_x>L_y$). The average ac magnetic field, $\overline{h}_z\equiv\frac{1}{L_xL_y}\int_0^{L_x}\int_0^{L_y}h_zdxdy,$ is
\begin{equation}\label{Eq4}
\overline{h}_z=h_z^{s1}Z_1+h_z^{s3}Z_2,
\end{equation}
where $$Z_1=8\sum_{m=1,3}^{\infty}\frac{\tanh(k_mL_y)}{\pi^2m^2k_mL_y}$$
and $$Z_2=8\sum_{m=1,3}^{\infty}\frac{\tanh(q_mL_x)}{\pi^2m^2q_mL_x}.$$ It was taken into account that due to the symmetry $h^{s1}=h^{s2}$ and $h^{s3}=h^{s4}$. The observed magnetic susceptibility is
\begin{equation}\label{Eq5}
\chi_1=(\overline{h}_z/h_0-1)/4\pi.
\end{equation}
Using these expressions and experimental data one can find the surface currents at each side of the sample. If $\overrightarrow{H_0}$ lays in the $XZ$-plane and the angle between dc field and $0Z$ axis is $\theta$
\begin{equation}\label{Eq5a}
1+4\pi\widetilde{\chi}_1=Z_1h_z^{s1}/h_0+Z_2h_z^{s2}/h_0,
\end{equation}
while if $\overrightarrow{H_0}$ lays in the $YZ$-plane and forms the same angle with $0Z$ axis, then
\begin{equation}\label{Eq5b}
1+4\pi\widetilde{\widetilde{\chi}}_{1}=Z_1h_z^{s2}/h_0+Z_2h_z^{s1}/h_0,
\end{equation}
where $\widetilde{\chi}_1$ and $\widetilde{\widetilde{\chi}}_1$ are susceptibilities for G1 and G2 configurations of the dc field, respectively.
 These two equations allow us to find both $h_z^{s1}$ and $h_z^{s2}$ from experimental data and to calculate the surface currents from Eq.~(\ref{Eq2}). The bulk conductivity in a normal state $\sigma$ can be found from the ac response at temperatures $T>T_c$. Under these conditions $h^{s1}=h^{s3}=h_0$. Fig.~\ref{f-3a} demonstrates the frequency dependence of $\chi_1^{\prime\prime}$ at T=7.5 K. This experiment was carried out on the setup described in Ref.~\cite{LEV2}. Mapping these data by Eq. (\ref{Eq5a}) we obtain $\sigma=8\times 10^{16}$~CGS. This value is in good agreement with the result of Ref.~\cite{JUNO} obtained for actually the same sample (it was cut from the same bulb).

\begin{figure}
     \begin{center}
    \leavevmode
 \includegraphics[width=0.9\linewidth]{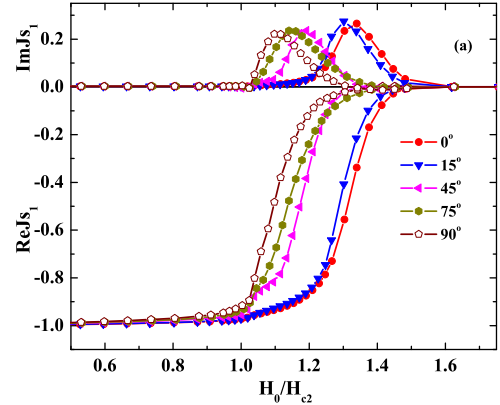}
\includegraphics[width=0.9\linewidth]{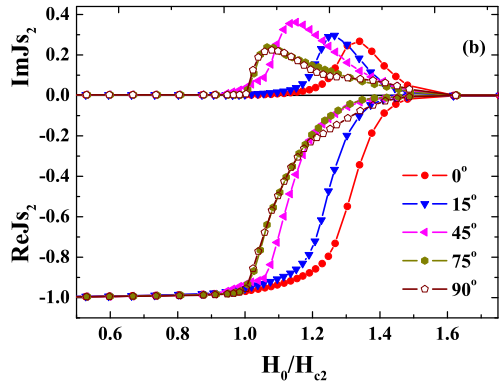}

\caption{(Color online) Panel (a): Field dependence of the surface current in the case when dc field makes angle with the surface.
Panel (b): Field dependence of the surface current when both dc and ac fields are parallel to the surface but make some angle with each other.}
     \label{f-3}
     \end{center}
     \end{figure}

\begin{figure}
     \begin{center}
    \leavevmode

\includegraphics[width=0.9\linewidth]{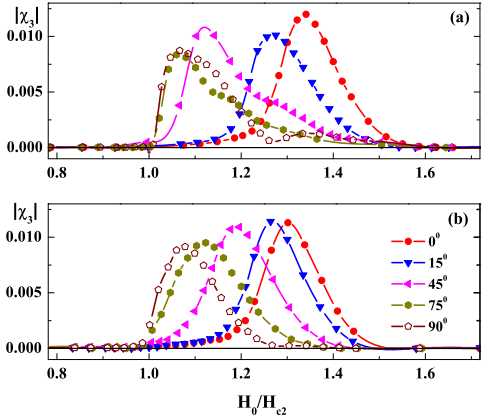}
\caption{(Color online) Field dependence of $|\chi_{3}|$ for different configurations of the dc field.
 Panel (a): the case when the dc field remains parallel to the \textit{YZ} sample surface but forms some angle with respect to the ac field.
Panel (b): the case when the dc field  remains parallel to the \textit{XZ} sample plane.}

     \label{f-3dd}
     \end{center}
     \end{figure}

\begin{figure}
     \begin{center}
    \leavevmode
 \includegraphics[width=0.9\linewidth]{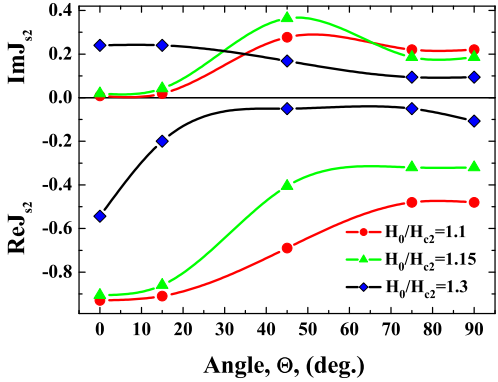}

\caption{(Color online)
Real and imaginary parts of the surface current as a function of the angle between ac and dc fields  for several dc fields. Both fields are parallel to the surface.}

     \label{f-5}
     \end{center}
     \end{figure}

Experimental results can be presented in terms of the surface current, $J_{s}$, normalized in such way that the complete shielding by the surface current corresponds to $J_{s}=-1$ while for the sample in a normal state $J_{s}=0$. Figs.~\ref{f-3}a and \ref{f-3}b shows $J_s$ as a function of the dc magnetic field for several angles. Fig.~\ref{f-3}a corresponds to the case when the dc field makes some angle with the surface while Fig.~\ref{f-3}b refers to the case when $\overrightarrow{H}_0$ and $\overrightarrow{h}_0$ are parallel to the surface but there is an angle between them.
We see that in the perpendicular dc field the superconducting transition at the surface takes place at $H_0>H_{c2}$. The observed response at the third harmonic confirms this conclusion also. The third harmonic vanishes if the dc field becomes smaller than $H_{c2}$. The nonlinear response disappears simultaneously with the disappearance of the surface current in an increased dc field. Moreover, the nonlinear response depends on the orientation the dc field, as is shown by Figs.~\ref{f-3dd}a and~\ref{f-3dd}b.
The orientation of the dc and ac magnetic fields here was the same as for Figs.~\ref{f-2}a and \ref{f-2}b. For third harmonic data we did not perform the procedure of separating the response from each sample side. However, the first harmonic data show that the wide sides provide the largest contribution to the ac response. This is why the dependence of $|\chi_3|$ on the angle shown in Fig.~\ref{f-3dd}a reflects the dependence of the nonlinear response by SSS on the angle between a parallel to the surface ac and dc fields. When $\overrightarrow{H}_0$ is parallel to the surface, the observed onset of the superconducting transition does not depend on the angle between the ac and dc fields as was expected, Fig.~\ref{f-3}b,  but the value of $J_s$ depends on this angle. Fig.~\ref{f-5} demonstrates the angular dependence of the surface current $J_s$ for several dc fields. This dependence is nonmonotonic at low dc fields. Detailed discussion of the angular dependence of $J_s$ will be reported in our forthcoming publications.

Existence of the SSS in the perpendicular dc field for $H_0>H_{c2}$ can be ascribed to the surface roughness. It is well known that the properties of the surface strongly affects the ac response of SSS and $H_{c3}$ depends on the surface roughness too~\cite{SWR,KAR,JUR}. SSS are localized in the surface layer with a thickness of a few coherence lengths. For a rough surface the dc field is not parallel to the local normal even in the perpendicular to the surface dc field and one could expect the manifestation of SSS in fields $H_0>H_{c2}$.
The value of $H_{c3}$ could be calculated from the first GL equation for a given magnetic field
\begin{equation}\label{Eq7}
\ln(T_c/T)\{-\Psi+|\Psi|^2\Psi\}+(i\overrightarrow{\nabla}+\overrightarrow{a})^2\Psi=0,
\end{equation}
with boundary conditions $\frac{d\Psi}{d\overrightarrow{n}}|_S=0$ at the surface. Here $\Psi=\Delta/\Delta_0$ is the
dimensionless order parameter, $\Delta_0$ is the order parameter at a given temperature and zero dc field, $\overrightarrow{a} =2e\overrightarrow{A}/\hbar c$, $\overrightarrow{A}$ is a vector potential ($\text{curl} \overrightarrow{A}=\overrightarrow{H_0}$), the length unit is $\xi_0$ - the coherence length at $T=0$, and $\overrightarrow{n}$ is a local outer normal to the sample surface. For a really rough surface this is a 3D equation and in order to simplify the problem we considered only the 2D version where the surface profile depended only on one direction (\textit{Z}-axis). Because the analytical solution is not possible for a rough surface profile we used the numerical method. Eq.(\ref{Eq7}) had been transformed to
\begin{eqnarray}
 \ln(T_c/T)\left\{-\Psi(n,m)+\left |\Psi\right |^2\Psi(n,m)\right\}-\nonumber\\
    \frac{\Psi(n+1,m)\Psi(n-1,m)-2\Psi(n,m)}{dx^2}-\nonumber\\
     \frac{\Psi(n+1,m)\Psi(n-1,m)-2\Psi(n,m)}{dz^2}+\nonumber\\
     i[\Psi(n,m+1)-\Psi(n,m-1)]mdzH_0+\nonumber\\
     H_0^2m^2dz^2\Psi(n,m)=\gamma\frac{d\Psi(n,m)}{dt}\nonumber
\end{eqnarray}
at the 2D-grid with steps $dx$ and $dz$.
The stationary nontrivial solutions of this set of equations was looked for, and the maximal $H_0$, for which such solution could be found, was considered as $H_{c3}$.
\begin{figure}
     \begin{center}
    \leavevmode
\includegraphics[width=0.9\linewidth]{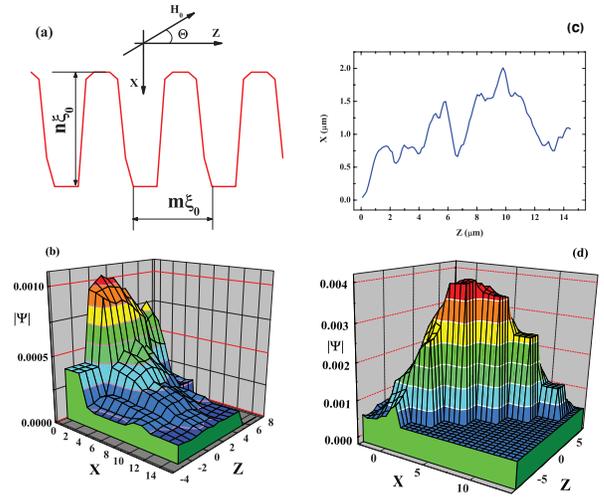}

\caption{(Color online)  Panel (a): The model profile of the rough surface and orientation of dc magnetic field. Panel (b): Calculated order parameter (absolute value) for SSS for the model surface in the dc field $H>H_{c2}$ and $\theta=45^{\circ}$). Panel (c): Measured by AFM actual profile of the surface.  Panel (d): The calculated order parameter near the surface for actual surface profile  ($\theta=45^{\circ}$). $\xi_0$ is unit length for (b) and (d) panels. See text.}

     \label{f-6}
     \end{center}
     \end{figure}
Actually, a grid with $10^4$ points was used. Calculations were performed for several types of the surface roughness, for a model with sinusoidal surface profile, Fig~\ref{f-6}a, and for the actual surface profile that was measured by an atomic force microscope (AFM), Fig.~\ref{f-6}c.

The shape of the superconducting nucleus near the rough surface is shown for illustration by Fig~\ref{f-6}b and Fig~\ref{f-6}d. Fig~\ref{f-6}b demonstrates the results of the numeric calculation of the spatial distribution of the order parameter near the surface with a sinusoidal roughness with period $m\times\xi_0$ and amplitude $n\times\xi_0$ while Fig.~\ref{f-6}d shows the order parameter for the surface profile measured by the AFM. The angle between the dc field direction and the sample surface equals 45$^{\circ}$ in both cases. Careful analysis of these pictures shows that a superconducting phase nucleates near such points where the angle between dc field and local normal is maximal, close to $\pi/2$.

\begin{figure}
     \begin{center}
    \leavevmode
\includegraphics[width=0.9\linewidth]{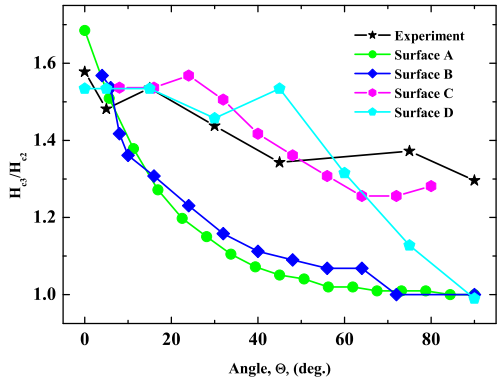}

\caption{(Color online) . The angular dependence of $H_{c3}/H_{c2}$ ratio: experiment and calculations for several models of the surface roughness. The curve $A$ corresponds to the ideally smooth surface; the  curve $B$ - to sinusoidal surface with $m=20$ and $n=1$; curve $C$ - to the model surface with $m=10$ and $n=5$; and curve $D$ - to the measured by AFM profile. See text.}

     \label{f-7}
     \end{center}
     \end{figure}

Fig.~\ref{f-7} shows both the experimental and calculated $H_{c3}/H_{c2}(\theta)$ dependence for several models of surface roughness. Curve ($A$) corresponds to the ideally smooth surface; ($B$) to a sinusoidal model with period $20\times\xi_0$ and amplitude $1\times\xi_0$; ($C$) - for period $10\times\xi_0$ and amplitude $5\times\xi_0$; and curve ($D$) correspond to the actual surface profile measured by AFM. The last curve presents the maximal value of $H_{c3}$ that was calculated at a given angle for profiles measured by the AFM. In case $B$, when the surface is relatively smooth $H_{c3}/H_{c2}=1.56$ for the parallel to the surface dc field. This is in good agreement with the experimental value 1.57. However, the calculated $H_{c3}$ decreases with an angle and for $\theta=\pi/2$ $H_{c3}/H_{c2}=1$ while the experimental value is 1.28 (Fig.~\ref{f-7}). For a more steep roughness $H_{c3}/H_{c2}\approx 1.25$ at angles close to $\pi/2$ (curve $C$). This ratio fits well the experimental value of 1.28 for a perpendicular dc field. However, the curve $C$ was calculated for a very steep roughness that was not observed for our sample by AFM. We have to note that the scale at this model rough surface is several tenths of nanometers.
One could expect the comparatively large $H_{c3}$ in the normal dc field if there are sufficient number of spots at the surface for which the angle between the dc field and the local normal is considerably large, but the AFM data do not confirm this.
We see that actual surface roughness could not yield such a large value of $H_{c3}/H_{c2}$ in a perpendicular dc field. Another hypothesis that there is a superconducting network in the bulk which is not seen by dc measurements~\cite{STR} at $H_0>H_{c2}$ leads to the requirement that the bulk has to have the normal conductivity that is approximately 100 times larger then was measured.

\section{Conclusions}

We have studied the low frequency ac response of a YB$_6$ single crystal in tilted dc magnetic fields. The developed approach has permitted us to obtain the value of the ac surface current at each side of the sample. The $H_{c3}$ angular dependence when dc field rotates out of the plane was obtained. The experiment showed that in the perpendicular to the surface dc field the transition to superconducting state took place in a field $H_0\approx 1.28H_{c2}$. This unexpected experimental result cannot be ascribed to the possible affect of the surface roughness. We demonstrated also that the induced ac surface current is a function of an angle between ac and dc fields when both fields are parallel to the surface.

\acknowledgments

This work was supported by the Israeli Ministry of Science (Israel -
Ukraine and Israel-Russia funds), and by the Klatchky foundation for superconductivity.
We wish to thank I. Felner, Ya. Kopelevich and A.E. Koshelev for fruitful discussions.
Help with programming provided by W. Glaberson is gratefully appreciated.

\end{document}